\documentclass[reprint,
 amsmath,amssymb,
 aps,
]{revtex4-2}

\usepackage{hyperref}
\usepackage{integrable,twistor}
\usepackage{dcolumn}
\usepackage{bm}

\begin{document}

\preprint{APS/123-QED}

\title{Towards celestial chiral algebras of self-dual black holes}

\author{Giuseppe Bogna}
\email{giuseppe.bogna@maths.ox.ac.uk}
\affiliation{The Mathematical Institute, University of Oxford, UK}

\author{Simon Heuveline}
\email{sph48@cam.ac.uk}
\affiliation{Dept. Applied Maths \& Theoretical Physics, University of Cambridge, UK}

\date{\today}

\begin{abstract}
In this note, we show that several self-dual spacetimes previously studied in the context of celestial and twisted holography arise as limits of a certain Taub-NUT AdS$_4$ metric, the Pedersen metric, in which their Mass, NUT charge and cosmological constant obey a self-duality relation. In particular, self-dual Taub-NUT,  a singular double cover of Eguchi-Hanson space, Euclidean AdS$_4$, and non-compact $\mathbb{CP}^2$, which is conformally equivalent to Burns space, arise as special limits of the Pedersen metric. The Pedersen metric can be derived from a curved twistor space which we conjecture to arise from a backreaction of self-dual gravity in the presence of a cosmological constant when coupled to a defect operator wrapping a certain $\mathbb{CP}^1$ at infinity. The curved twistor space gives rise to a $2$-parameter deformation of the celestial symmetry algebra $Lw_\wedge$ which reduces to previously studied algebras in various limits. %A companion paper will discuss additional details and relations to previously studied self-dual black holes.
\end{abstract}

\maketitle

\section{Introduction}

Celestial holography suggests, among other things, that collinear singularities of graviton scattering amplitudes are described by the OPE of a putative dual CFT \cite{Strominger:2017zoo, Raclariu:2021zjz, Pasterski:2021raf}. One of the great successes has been the insight that this duality is true at tree level which led to the discovery of new infinite dimensional symmetry algebras of tree-level amplitudes in flat space closely related to $w_{1+\infty}$ \cite{Strominger:2021mtt, Guevara:2021abz, Adamo:2021lrv}. 
Non-trivial backgrounds have been studied in the celestial context for several reasons \cite{Bittleston:2023bzp, Bittleston:2024rqe, Taylor:2023ajd, Bu:2022iak, Melton:2022fsf, Costello:2022wso, Mago:2021wje,  CarrilloGonzalez:2024sto, Lipstein:2023pih, McLoughlin:2024ldp, Adamo:2023zeh, Adamo:2024mqn, Costello:2022jpg, Costello:2023hmi, Garner:2023izn, Garner:2024tis, Gonzo:2022tjm, Melton:2023bjw, Melton:2023dee, Casali:2022fro, Fan:2022vbz, Kmec:2024nmu, Adamo:2023fbj, Crawley:2023brz} and it is known that considering (self-dual) gravity on such backgrounds can lead to deformations of these celestial symmetries \cite{Bittleston:2023bzp, Bittleston:2024rqe, Taylor:2023ajd, Bu:2022iak, Melton:2022fsf, Costello:2022wso, Mago:2021wje, CarrilloGonzalez:2024sto} or the absence thereof \cite{Adamo:2023fbj}. Deformed algebras were previously identified, among other backgrounds, for Eguchi-Hanson space \cite{Bittleston:2023bzp} and AdS$_4$ \cite{Bittleston:2024rqe, Taylor:2023ajd, Lipstein:2023pih}. In this work, we will derive a $2$-parameter deformation of $Lw_\wedge$ which interpolates between these two deformations and provide its bulk interpretation.
 
We consider a backreaction from coupling self-dual gravity in the presence of a cosmological constant to a defect operator on twistor space. This defect wraps $\mathbb{CP}^1$ over a spacetime point at infinity, which backreacts to the twistor space of self-dual Taub-NUT when $\Lambda=0$.
An interpretation of such a defect in the light of \cite{Crawley:2023brz} will be provided in future work. 
When $\Lambda< 0$, we give strong evidence that the resulting backreaction leads to a self-dual limit of Wick-rotated Taub-NUT AdS$_4$ black holes, which we refer to as the Pedersen metric. In particular, 
 the Pedersen metric can be viewed as a $\Lambda\neq 0$ generalization of the self-dual black holes recently considered in \cite{Crawley:2021auj, Crawley:2023brz, Guevara:2023wlr, Easson:2023ytf, Desai:2024fgr}. There, an important role is played by working in Kleinian signature but here, we will only work in Euclidean signature and with a negative cosmological constant. In the near future, we will further discuss properties of the Pedersen metric in Kleinian signature, $\Lambda>0$, which can both be obtained from the same twistor space we consider here, and further relations to \cite{Crawley:2021auj, Crawley:2023brz} such as the role of non-vanishing angular momentum in the presence of a cosmological constant. 

 The Pedersen metric depends on two parameters and further relations can be imposed between these in which it reduces to previously studied self-dual geometries such as self-dual Taub-NUT,  a singular double cover of Eguchi-Hanson space, Euclidean AdS$_4$, and non-compact $\mathbb{CP}^2$, which is conformally equivalent to Burns space. From its $2$-parameter twistor space we derive a conjectured $2$-parameter deformation of $Lw_\wedge=L\mathfrak{ham}(\mathbb{C}^2/\mathbb{Z}_2)$ (following the conventions of \cite{Bittleston:2023bzp, Bittleston:2022jeq}) which reduces to the expected deformations in various limits.  

Here, we only consider tree-level results with the hope that the algebra we find might be extended beyond tree-level by means of some top-down construction. 
In the case of Burns space, a fully-fledged holographically dual chiral algebra was identified \cite{Costello:2023hmi, Costello:2022jpg} but the gravitational part of the algebra is much larger than the previously discussed symmetries 
\cite{Strominger:2021mtt} since the $4$-dimensional bulk gravitational theory is a self-dual subsector of conformal gravity, Mabuchi gravity, rather than Einstein gravity.
The Pedersen metric is conformally equivalent to a $2$-parameter family of scalar flat Kähler manifolds i.e. solutions to the classical field equations of Mabuchi gravity \cite{Costello:2022jpg, Costello:2021bah} which interpolates between Burns space and a singular double cover of Eguchi-Hanson space. So, we expect to obtain the Pedersen twistor space in the context of the recent top-down construction \cite{Bittleston:2024efo} or \cite{Costello:2023hmi, Costello:2022jpg}. From this, we hope to obtain amplitudes on a Pedersen background from some chiral algebra computation which could be matched with results in the literature \cite{Zoubos:2002cw}.
Since a non-trivial thermodynamic phase structure is known to exist for the Pedersen metric \cite{Chamblin:1998pz, Hawking:1998ct, Martelli:2012sz} such considerations might lead to new insights into the non-perturbative structure of twisted holography. 

\clearpage

Throughout this paper we make free use of spinor conventions $\la\lambda\kappa\ra = \lambda^\al\kappa_\al=\epsilon^{\al\beta}\lambda_\beta\kappa_\al$ and similarly $[\tilde\lambda\tilde\kappa]=\tilde\lambda^\da\tilde\kappa_\da$. The twistor space $\PT$ of flat Minkowski space is the total space of $\cO(1)\oplus\cO(1)\to\CP^1 \cong \mathbb{CP}^3\setminus \mathbb{CP}^1$. We will use homogeneous coordinates $\lambda_\alpha$ on the $\CP^1$ base, and $\mu^\da$ on the fibres, collectively denoting these by $Z^{A}=(\mu^\da,\lambda_\al)\in \mathbb{CP}^3$ such that $\lambda_\alpha \neq 0$. Euclidean spinor conjugation is given by $\omega^{\alpha}=(a,b)\mapsto \hat{\omega}^{\alpha}=(-\bar{b},\bar{a})$ and $\kappa^{\dot{\alpha}}=(c,d)\mapsto \hat{\kappa}^{\dot{\alpha}}=(-\bar{d},\bar{c})$.

%%%%%%%%%%%%%%%%%%%%%%%%%%%%%%%%%%%%%%%%%%%

\section{The Pedersen metric and its limits}
\subsection{Self-dual Pleba\'{n}ski-Demia\'{n}ski Spacetimes
}
\label{subsec:sdPlebanski}

A well-known $7$-parameter generalization of the Kerr metric was constructed by Pleba\'{n}ski-Demia\'{n}ski in \cite{Plebanski:1976gy}. We will not be interested in acceleration or electromagnetic charges, so we set the corresponding parameters to $0$. The resulting $4$-parameter family of rotating Taub-NUT-AdS black holes reads \cite{Griffiths:2005qp, Rodriguez:2021hks}
\bea
\label{eq:TNAdS4}
\dif s^2=& -\frac{\Delta}{\Sigma}\left(\dif t+(2n\cos\theta-a\sin^2\theta)\frac{\dif\phi}{\Xi}\right)^2\\
&+\frac{\Delta_\theta}{\Sigma}\left(a\,\dif t-(r^2+a^2+n^2)\frac{\dif \phi}{\Xi}\right)^2\\
&+\frac{\Sigma}{\Delta}\,\dif r^2+\frac{\Sigma}{\Delta_\theta}\sin^2\theta\,\dif \theta^2\,,
\eea
where
\bea
\Sigma&=r^2+(n+a\,\cos \theta)^2\,,\\
\frac{\Delta_\theta}{\sin^2\theta}&=1-\frac{4an \cos \theta}{l^2}-\frac{a^2\cos^2 \theta}{l^2}\,,\\
\Delta&=r^2+a^2-2mr-n^2\\
&\quad+\frac{3(a^2-n^2)\,n^2+(a^2+6n^2)\,r^2+r^4}{l^2}\,,\\
\Xi&=1-\frac{a^2}{l^2}\,.
\eea
The metric \eqref{eq:TNAdS4} is known to solve Einstein's equations in the presence of a cosmological constant $\Lambda=-3/l^2$. The remaining three parameters $(m,n,a)$ are related to the mass, the NUT charge and the angular momentum of the black hole. 

We will consider the case $a=0$. In the absence of $\Lambda$, i.e. in the limit $l\rightarrow \infty$, the metric \eqref{eq:TNAdS4} has a self-dual limit $n=-\im M$, $m=M$, in which it can be analytically continued to give a Euclidean self-dual metric commonly referred to as self-dual Taub-NUT \cite{Hawking:1976jb}. Alternatively, it can be continued to Kleinian signature \cite{Crawley:2021auj, Easson:2023ytf}, where the metric has a genuine horizon beyond which it can be extended with the maximal extension having a singularity at $r=-M$. This justifies the previously used terminology \emph{self-dual black hole}. It was also argued in \cite{Crawley:2021auj, Desai:2024fgr} that the parameter $a$ can be reintroduced with a diffeomorphism closely related to the Newman-Janis shift and it is a natural question if this can be extended to $\Lambda \neq 0$.

In the presence of a non-vanishing cosmological constant, $\Lambda\neq0$, there exists a similar self-dual limit, which can be obtained from computing the components of the Weyl-tensor in the Newman-Penrose formalism using a certain null-tetrad \cite{Griffiths:2005qp}. The only non-vanishing component of the anti-self-dual Weyl tensor is
\begin{equation}
    \psi_2=\frac{4n^3+l^2(-\im m+n)}{l^2(n-\im r)^3}\,,
\end{equation}
together with its complex conjugate $\tilde\psi_2$. We can set either of them to zero by imposing \be
n=\pm\im M\,,\qquad m=M\left(1-\frac{4M^2}{l^2}\right)\,,\label{eq:self-dualpoint}
\ee
which gives an (anti-)self-dual metric that has previously been studied in various contexts \cite{Gibbons:1978zy, Zoubos:2002cw, Chamblin:1998pz, Ivanov:1998de}. 
The resulting metric can be brought into triaxial form with a suitable diffeomorphism that we discuss in section \ref{sec:Diffeos}. This form of the metric was first constructed by Pedersen \cite{Pedersen:1986vup, pedersen1985geometry}, so we will refer to it as the \emph{Pedersen metric}.

We also note that when $a\neq0$ the (anti)-self-duality condition is further deformed to
\be
n=\pm\im M\,, \qquad m=M\left(1-\frac{a^2+4M^2}{l^2}\right)\,,
\ee
which will be further discussed in the near future.

\subsection{The Pedersen metric}
\label{sec:Pedersenmetric}
The triaxial form of the Euclidean Pedersen metric on the $4$-dimensional ball of radius $l$ is
\be
\label{eq:PedMet}
\d s^2=f^2(r)\left(h_r(r)\d r^2+h_{12}(r)(\sigma_1^2+\sigma_2^2)+h_{3}(r)\sigma_3^2\right)\,.
\ee
where
\bea
    f(r)&=\frac{2}{1-r^2/l^2}\,,\qquad h_r(r)=\frac{1+\nu^2r^2}{1+\nu^2r^4/l^2}\,,\\ h_{12}(r)&=r^2(1+\nu^2r^2)\,,\qquad h_3(r)=\frac{r^2}{h_r(r)}\,,
\eea
and $\sigma_i$ denote the standard $\SU(2)$ left-invariant 1-forms. We can see that the conformal boundary of the spacetime is a biaxially squashed 3-sphere known as Berger's sphere: setting $r=l$ in \eqref{eq:PedMet}, we find the boundary metric
\begin{equation}
\label{eq:oblatesquashedS3}
    \d s^2_{3}=\sigma_1^2+\sigma_2^2+\frac{1}{1+\nu^2l^2}\sigma_3^2\,,
\end{equation}
up to an overall constant. Since $(1+\nu^2l^2)^{-1}\leq 1$, the metric \eqref{eq:oblatesquashedS3} is an oblate squashing of the $3$-sphere. For this reason, we will also refer to the metric in \eqref{eq:PedMet} as the \emph{oblate Pedersen metric}. The spacetime metric represents an explicit realization of a theorem by LeBrun \cite{LeBrun:1982vjh}, where it is shown that any 3-manifold with a Riemannian conformal structure is the conformal infinity of a self-dual Riemannian 4-manifold satisfying Einstein's equations with a negative cosmological constant. The theorem ensures the existence of the bulk 4-manifold only in a collared neighbourhood of its conformal boundary, but the Pedersen metric is a complete metric inside the entire $4$-ball with radius $l$. In the language of Lebrun \cite{lebrun1991complete}, this means that the biaxially squashed $3$-sphere is of positive frequency (with the appropriate orientation). More generally, Hitchin showed that that every left-invariant conformal structure on $S^3$ has positive frequency leading to a generalisation of the Pedersen metric \cite{Hitchin:1995hxv}.

\subsection{Relation to Mabuchi gravity}
As shown in the supplemental material section \ref{sec:Diffeos}, the Pedersen metric is diffeomorphic to
\begin{equation}
    \d s^2=\frac{\nu l^3}{(\sin\chi-\nu l\cos\chi)^2}\left(V^{-1}(\d\tau+A)^2+V\d \Omega^2_3\right)\,,\label{eq:s3_gibbons-hawking}
\end{equation}
where we introduced the Higgs field and gauge potential on the 3-sphere
\begin{equation}
    V=\nu l+\cot\chi\,,\qquad A=-\cos\theta\,\d\phi\,.
\end{equation}
$\d \Omega_3^2 =\d\chi^2+\sin^2\chi\,\d\Omega_2^2$ denotes the canonical metric on the 3-sphere. It is straightforward to show that the pair $(V,A)$ satisfies the Bogomolny equation $\star_3\,\d V=\d A$, where $\star_3$ denotes the Hodge star operator on $S^3$ with respect to $\d\Omega_3^2$.

It is also possible to analytically continue the metric \eqref{eq:PedMet} by $\nu \mapsto \im \nu$. Provided that $\nu l < 1$, the metric is still complete on the open ball $r<l$. We will refer to this as the \emph{prolate Pedersen metric}, because its boundary conformal structure is a prolate squashing of $S^3$ in contrast to the oblate Pedersen metric \eqref{eq:PedMet}. The prolate Pedersen metric can also be written as
\be
\label{eq:eq:h3_gibbons-hawking}
\d s^2=\frac{\nu l^3}{(\sinh\chi-\nu l\cosh\chi)^2}(V_H^{-1}(\d\tau+A_H)^2+V_H\d s_{\mathbb{H}^3}^2)\,,
\ee
where now the pair $(V_H,A_H)$ is defined on $\mathbb{H}^3$
\begin{equation}
    V_H=-\nu l+\coth\chi\,,\qquad A_H=-\cos\theta\,\d\phi\,,
\end{equation}
and $\d s_{\mathbb{H}^3}^2=\d\chi^2+\sinh^2\chi\,\d\Omega_2^2$ denotes the standard metric on $\mathbb{H}^3$.
The Pedersen metric can thus be understood as arising from a generalized Gibbons-Hawking ansatz \cite{gibbons1978gravitational} where one replaces the $\R^3$ with $S^3$ or $\mathbb{H}^3$, respectively in the oblate and prolate case \cite{pedersen1985geometry, Pedersen:1986vup}.

A conformal factor can be included to turn \eqref{eq:eq:h3_gibbons-hawking} into
\be
\label{eq:ProlPedMetconf}
\d s^2=\nu l^3q(\chi,\theta, \phi)^2(V_H^{-1}(\d\tau+A_H)^2+V_H\d s_{\mathbb{H}^3}^2)\,,
\ee
which has been shown to be scalar flat Kähler by Lebrun \cite{lebrun1991explicit}. $q(\chi,\theta, \phi)$ is a horospherical height function \cite{lebrun1991complete}, which is explicitly given by the coordinate $q$ after changing to half-plane coordinates $\dif s_{\mathbb{H}^3}^2=q^{-2}(\dif x^2+\dif y^2 +\dif q^2)$ on the $\mathbb{H}^3$ factor \cite{costa2001description}. Since twistor spaces are a priori agnostic about conformal factors, we see that the Pedersen twistor space also leads to a two-parameter family of solutions to the equations of motion of Mabuchi gravity \cite{Costello:2023hmi, Costello:2021bah}. In the next section and \ref{sec:Diffeos}, we will discuss that this class of metrics particularly interpolates between Burns space and a double-cover of Eguchi-Hanson space leading to a conjectured relation to \cite{Bittleston:2024efo}.

\subsection{Various limits}

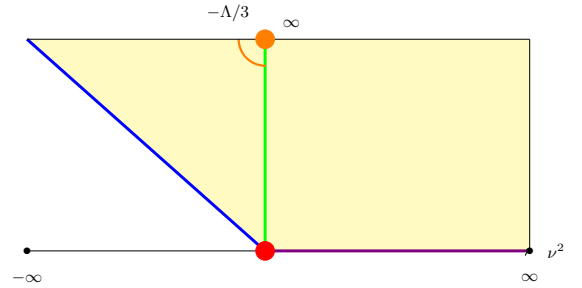
\begin{figure}[h!]
	\centering
 	\scalebox{.7}{\begin{tikzpicture}[scale=1]
%\draw[step=0.5,black,thin] (0,0) grid (10,5);
\draw[black] (0.5,0) -- (5,0);
\draw[->] (5,0) -- (10,0);
\draw[->] (5,0) -- (5,4);

\fill[yellow!30] (5,0)--(5,4)--(0.5,4);
\fill[yellow!30] (5,0)--(5,4)--(10,4)--(10,0);

\draw[green, ultra thick] (5,0) -- (5,4);
\draw[violet, ultra thick] (5,0) -- (10,0);
\draw[blue, ultra thick] (5,0) -- (0.5,4); %.. controls (4,2.8) and (3,3.5) .. (0.5,4);
\begin{scope}
    \clip (4,3) rectangle (5,4);
    \draw[orange, very thick] (5,4) circle(0.5);
\end{scope}

\draw (0.5,4) -- (5,4);
\draw (5,4) -- (10,4);
\draw (10,4) -- (10,0);

\node at (10.5,0){$\nu^2$};
\node at (4.3,4.5){$-\Lambda/3$};
\node at (10,0){$\bullet$};
\node at (10,-0.5){$\infty$};
\node at (5,4){$\bullet$};
\node at (5.5,4.3){$\infty$};
\node at (0.5,0){$\bullet$};
\node at (0.5,-0.5){$-\infty$};

\filldraw[red] (5,0) circle (5pt);
\filldraw[orange] (5,4) circle (5pt);

\end{tikzpicture}}
\caption{the Pedersen metric (yellow) has various limits: a singular double cover of Eguchi-Hanson space (orange), self-dual Taub-NUT (violet), $\widetilde{\mathbb{CP}}^2$ (blue), Euclidean AdS$_4$ (green), and $\mathbb{R}^4$ (red). $\nu^2<0$ makes up the prolate case while $\nu^2>0$ makes up the oblate case.} 
\label{Pedersen}
\label{fig:Ped}
\end{figure}

The oblate Pedersen metric \eqref{eq:PedMet} and its continuation to the prolate Pedersen metric \eqref{eq:eq:h3_gibbons-hawking} depend on two parameters $\nu^2$ and $\Lambda=-3/l^2$. Note that $\nu$ in \eqref{eq:PedMet} is either real with any  $l>0$ (oblate case) or purely imaginary with $-\im \nu l<1$ (prolate case) in order to ensure completeness. The resulting region is displayed in yellow in figure \ref{fig:Ped}. There are various limiting cases in which the Pedersen metric turns into spacetimes that were previously studied in the context of celestial and twisted holography \cite{Crawley:2021auj, Crawley:2023brz, Bittleston:2023bzp, Bittleston:2024rqe, Costello:2023hmi, Costello:2022jpg, Bittleston:2024efo}. These limits are displayed in figure \ref{fig:Ped} and discussed in the supplemental material section \ref{sec:Diffeos}. 

%%%%%%%%%%%%%%%%%%%%%%%%%%%%%%%%%%%%%%%%%%%

\section{Self-dual gravity on twistor space}
\label{sec:sec3}
We will consider the twistor uplift of self-dual gravity in the presence of a cosmological constant described by the Mason-Wolf action \cite{Mason:2007ct}
\begin{equation}
    \label{eq:twistor-defect-action}
    S_\Lambda [g,h] =  \int_{\PT_\Lambda}\Dif^3Z\wedge g\wedge \left(\bar\p h + \frac{1}{2}\{h,h\}_\Lambda\right) \,,
\end{equation}
where $g\in \Omega^{0,1}(\PT_\Lambda,\mathcal{O}(-6))$ and $h\in \Omega^{0,1}(\PT_\Lambda,\mathcal{O}(-2))$. The holomorphic $\mathcal{O}(-2)$-valued Jacobi bracket $\{\,\,,\,\,\}_\Lambda$ is defined through an infinity twistor \cite{Ward:1980am, Penrose:1976jq} and we choose to work in the following homogeneous coordinates
\bea
\label{eq:LambdaBrack}
&\{f,g\}_\Lambda=\varepsilon^{\dot\alpha\dot\beta}\frac{\p f}{\p\mu^{\dot\alpha}}\frac{\p g}{\p\mu^{\dot\beta}}+\Lambda\,\varepsilon_{\alpha\beta}\frac{\p f}{\p\lambda_\alpha}\frac{\p g}{\p\lambda_\beta}\,.
\eea
For a recent discussion of this action, we refer to \cite{Bittleston:2024rqe}. Note that working with homogeneous coordinates means that we are formally working on non-projective twistor, where \eqref{eq:LambdaBrack} is a (twisted) Poisson bracket whereas it becomes a (twisted) Jacobi bracket on projective twistor space. Below, we will formally work on a generalized non-projective twistor space of a curved spacetime which is formalised by the total space of the Swann-bundle over the curved twistor space \cite{Adamo:2021bej, swann1990hyperkahler}.

\subsection{Deforming twistor space with a defect operator}
\label{sec:Backreaction}

In \cite{Bittleston:2023bzp}, it was shown that the twistor space of Eguchi-Hanson space arises as a backreaction from including a defect operator wrapping $\mathbb{CP}^1_{\mu^{\dot{\alpha}}=0}$ in the flat twistor space of $\mathbb{C}^2/\mathbb{Z}_2$. We will now formally include a defect operator wrapping the twistor line $\mathbb{CP}^1_{\lambda_\alpha=0}$ in $\mathbb{CP}^3\supset \mathbb{PT}$ that couples electrically to $g$. Doing this can equivalently be viewed as introducing a boundary condition for $h$, a perspective which relates our work to \cite{Crawley:2023brz} and will be elaborated on in the future. 
\paragraph{Self-dual Taub-NUT}
\label{para:SDTN}
As a warm-up, let us consider the special case $\Lambda=0$ in which the backreacted twistor space is that of self-dual Taub-NUT. The deformed action reads
\bea
    \label{eq:twistor-defect-action1}
    S[g,h] &=  S_{0}[g,h]- \frac{\pi^2}{4M} \int_{\CP^1_{\lambda_\alpha=0}}\la\lambda\,\dif\lambda\ra\wedge\eta^2\,  g \,,
\eea
where  $\eta=\mu^{\dot{0}}\mu^{\dot{1}}=[\iota \mu][\mu\hat{\iota}]$ singles out a choice of dotted 
reference spinors $\iota, \hat{\iota}$ which breaks Lorentz invariance. Below, we will also use  $\mu^{+}=\mu^{\dot{0}}$ and $\mu^{-}=\mu^{\dot{1}}$ for notational convenience. The latter term in \eqref{eq:twistor-defect-action1} describes the coupling to the defect. $1/M$ is to be viewed as a coupling constant controlling the strength of the coupling. For $1/M\neq 0$, we can vary $g$ to obtain the deformed equation of motion for $h$
\begin{equation}
    \dbar h+\frac{1}{2}\{h,h\}_0=\frac{\pi^2\eta^2}{2M}\bar\delta^{(2)}(\lambda)\,.
\end{equation}
The equation for $g$ remains unchanged. Similar to \cite{Bittleston:2023bzp}, we can solve the sourced equation by 
\begin{equation}
    h=\frac{\eta^2}{8M}\bar e^0\,,\qquad\bar e^0=\frac{\langle\hat\lambda\d\hat\lambda\ra}{\la\lambda\hat\lambda
    \ra^2}\,.\label{eq:sdtn_hamiltonian}
\end{equation}
since we have $\{h,h\}_0=0$, $\bar \partial h=0$ for $\la\lambda \hat{\lambda}\ra\neq 0$ and the correct normalization.
Note that up until this point, the backreaction is directly related to the one performed in \cite{Bittleston:2023bzp} by exchanging $\lambda_\alpha$ and $\mu^{\dot{\alpha}}$, which corresponds to a conformal inversion
\be
x^{\alpha\dal}\mapsto\frac{2 x^{\alpha\dal}}{x^2}
\ee
on spacetime, since
\be
\mu^\da =x^{\alpha \da}\lambda_\alpha \quad \iff \quad \frac{2 x_{\alpha \dal}}{x^2}\mu^\dal=\lambda_\alpha\,.
\ee
However, the Beltrami differential will not simply arise by exchanging $\lambda_\alpha$ and $\mu^{\dot{\alpha}}$ since the Poisson structure, or Jacobi structure when $\Lambda \neq 0$, of equation \eqref{eq:LambdaBrack} is not conformally invariant. We can see this explicitly by considering the deformed Dolbeaut operator which now reads
\begin{equation}
    \begin{aligned}
        \bar\nabla_0&=\dbar+\{h,\,\,\}_0\\&=\dbar +\frac{\eta}{4M}\bar e^0\wedge(\mu^{\dot 1}\mathcal{L}_{\dot 1}-\mu^{\dot 0}\mathcal{L}_{\dot 0})\,,
    \end{aligned}
\end{equation}
where $\mathcal{L}_{\dot\alpha}=\mathcal{L}_{\p/\p\mu^{\dot\alpha}}$ and $\mathcal{L}^\alpha=\mathcal{L}_{\p/\p\lambda_\alpha}$ denote the Lie derivatives \footnote{We would like to thank David Skinner for pointing this backreaction out to us.}. The coordinates $\lambda_\alpha$ and $\eta$ are still holomorphic in the deformed complex structure, but $\mu^\da$ are not. We can construct patchwise  (over $\mathbb{CP}^1$) holomorphic coordinates
\bea
\label{eq:rhopm}
\rho^\pm = \mu^{\pm}\,\exp\bigg(\pm\frac{\eta f(\lambda)}{4M}\bigg)\,,
\eea
where we introduced the patchwise-defined function
\be
\label{eq:f(lambda)}
f(\lambda) = \begin{cases}
    \displaystyle \frac{1}{\la\lambda \hat{\lambda}\ra}\frac{\hat\lambda_0}{\lambda_0}\qquad &\lambda_0\neq0\\
    \vspace{-1.2em}\\
    \displaystyle\frac{1}{\la\lambda \hat{\lambda}\ra}\frac{\hat\lambda_1}{\lambda_1}\qquad &\lambda_1\neq0
\end{cases}\;.
\ee
$\rho^\pm$ are holomorphic as a consequence of $\dbar f=\bar e^0$. 

On the overlap, $\rho^\pm$ patch according to the transition function 
\bea
\label{eq:TransitionSDTN}
\rho^\pm&\mapsto\rho^\pm\,\exp\bigg(\mp \frac{\eta \hat{\lambda}_0}{4M\la\lambda \hat{\lambda}\ra \lambda_0} \bigg)\exp\bigg(\pm \frac{\eta \hat{\lambda}_1}{4M\la\lambda \hat{\lambda}\ra \lambda_1} \bigg)  \\
&=\rho^{\pm}\, \exp\bigg(\pm \frac{\eta}{4M\lambda_0 \lambda_1} \bigg)\,.
\eea
Moreover, they also satisfy $\rho^+\rho^-=\eta$ on each patch. This backreacted geometry precisely matches Hitchin's description of the twistor space of self-dual Taub-NUT \cite{Hitchin:1979rts, hitchin1992hyper, Besse:1987pua}. Namely, it is defined as $\rho^+ \rho^-=\eta$ inside the total space of a sum of two line bundles $L^{1/4M}(1)$ and $L^{-1/4M}(1)$ over the minitwistor space \cite{calderbank2001einstein} of $\mathbb{R}^3$, $\mathbb{MT}=\mathcal{O}(2)\rightarrow \mathbb{CP}^1$
\bea
\label{eq:TwistorofSDTN}
&\{(\rho^+, \rho^-)\in L^{1/4M}(1)\oplus L^{-1/4M}(1): \rho^+ \rho^-=\eta\}\\
&\subset \text{Tot}(L^{1/4M}(1)\oplus L^{-1/4M}(1)\rightarrow \mathbb{MT})\,. 
\eea
The spacetime metric of self-dual Taub-NUT can be explicitly derived from this twistor space \cite{Besse:1987pua}. Note that the holomorphic coordinates \eqref{eq:rhopm} are formally equivalent to coordinates which previously appeared in the twisted holography context in \cite{Budzik:2023nnx}.

\paragraph{The Pedersen twistor space}
\label{para:PedTwistor}
Let us now consider the same backreaction in the presence of a cosmological constant.
Consider the action
\bea
    \label{eq:twistor-defect-action2}
    S[g,h] &=  S_{\Lambda}[g,h]- \frac{\pi^2\nu^2}{2} \int_{\CP^1_{\lambda_\alpha=0}}\la\lambda\,\dif\lambda\ra\wedge\eta^2\,  g \,,
\eea
The solution to the equations of motion is still given by 
\begin{equation}
    h=\frac{\nu^2\eta^2}{4}\bar e^0\,,
\end{equation}
since $\{h,h\}_\Lambda=0$. However, the presence of the new term in $\{\,\,,\,\,\}_\Lambda$ changes the Beltrami differential to
\begin{equation}
\label{eq:NablaQTN}
    \begin{aligned}
        \bar\nabla_\Lambda&=\dbar+\{h,-\}_{\Lambda}\\
        &=\dbar +\frac{\nu^2 \eta}{2}\bar e^0\wedge(\mu^{\dot 1}\mathcal{L}_{\dot 1}-\mu^{\dot 0}\mathcal{L}_{\dot 0})+\frac{{\Lambda}\nu^2\eta^2}{2}\frac{\hat\lambda_\alpha}{\la \lambda\,\hat\lambda\ra}\bar e^0\wedge\mathcal{L}^\alpha\,,
    \end{aligned}
\end{equation}
where we used the identity $\mathcal{L}^\alpha\bar e^0=2\hat\lambda^\alpha\bar e^0/\la\lambda\,\hat\lambda\ra$.
The coordinate $\eta=\mu^{\dot{0}}\mu^{\dot{1}}$ is still holomorphic but now neither $\mu^{\dot \alpha}$ nor $\lambda_\alpha$ are. Similar to \cite{Bittleston:2022nfr}, the $\lambda_\alpha$ coordinates get deformed to 
\begin{equation}
\label{eq:XYZ}
Y^{\alpha\beta}=\lambda^\alpha\lambda^\beta-\frac{\Lambda\nu^2 \eta^2}{2}\frac{\hat\lambda^\alpha\hat\lambda^\beta}{\la\lambda\,\hat\lambda\ra^2}\,.
\end{equation} 
which will be shown to obey $\bar{\nabla}_\Lambda Y^{\alpha \beta}=0$ in section \ref{sec:Holom}. Since the three holomorphic coordinates 
\begin{equation}
    X= Y^{11}\,,\quad Y= Y^{00}\,,\quad Z=-Y^{01}\,,  
\end{equation}
and $\eta$ all scale with weight $2$, and obey the  relation $Y^{\alpha \beta}Y_{\alpha \beta}=-\Lambda \nu^2\eta^2$, they form a quadric inside $\mathbb{CP}^3$
\be
Q_{\Lambda\nu^2}=\{(X,Y,Z,\eta)\in \mathbb{CP}^3: XY-Z^2+\frac{\Lambda \nu^2}{2} \eta^2=0\}\,.
\ee
which can be identified with $\mathbb{CP}^1\times \mathbb{CP}^1$, the twistor space of $S^3$ \cite{calderbank2001einstein}. It can be checked that the coordinates
\be
\label{eq:rhocoord}
\phi^{\pm}=\mu^\pm \Bigg(\frac{\sqrt{\tfrac{2}{\Lambda\nu^2}}+\eta f(\lambda)}{\sqrt{\tfrac{2}{\Lambda\nu^2}}-\eta f(\lambda)}\Bigg)^{\pm \sqrt{ \tfrac{\nu^2}{8\Lambda} }}\,.
\ee
are holomorphic. \eqref{eq:rhocoord} are defined on patches  $\lambda_0\neq 0$ and $\lambda_1 \neq 1$, which are somewhat unnatural in the given context. However, working with these patches gives
\bea
\label{eq:TransitionFunctionPhi}
& \Bigg(\frac{\sqrt{\tfrac{4M}{\Lambda}}+\eta \frac{\hat{\lambda}_0}{\lambda_0\langle\lambda\,\hat\lambda\rangle}}{\sqrt{\tfrac{4M}{\Lambda}}-\eta \frac{\hat{\lambda}_0}{\lambda_0\langle\lambda\,\hat\lambda\rangle}}\Bigg)^{\mp \tfrac{1}{2\sqrt{4\Lambda M}}}\Bigg(\frac{\sqrt{\tfrac{4M}{\Lambda}}+\eta \frac{\hat{\lambda}_1}{\lambda_1\langle\lambda\,\hat\lambda\rangle}}{\sqrt{\tfrac{4M}{\Lambda}}-\eta \frac{\hat{\lambda}_1}{\lambda_1\langle\lambda\,\hat\lambda\rangle}}\Bigg)^{\pm \tfrac{1}{2\sqrt{4\Lambda M}}}\\
&=\Bigg(\frac{Z+\sqrt{\tfrac{\Lambda}{4M}}\eta}{Z-\sqrt{\tfrac{\Lambda}{4M}}\eta}\Bigg)^{\pm \tfrac{1}{2\sqrt{4\Lambda M}}}\,,
\eea
which precisely matches with the expression found by Pedersen in equation (8.15) of \cite{pedersen1985geometry} when describing the twistor space of the Pedersen metric.  The Pedersen twistor space \cite{pedersen1985geometry,Pedersen:1986vup} is in fact given by a line bundle over $Q_{\Lambda\nu^2}$ with a transition function which is closely related to \eqref{eq:TransitionFunctionPhi}. We conjecture that our deformed twistor space described by the Beltrami differential \eqref{eq:NablaQTN} actually agrees with the full Pedersen twistor space, given by this line bundle. We hope to verify this conjecture more directly in the future by working with deformed holomorphic coordinates that are better adapted to the minitwistor space of $S^3$ or $\mathbb{H}^3$.

Further evidence for this conjecture is given by the fact that our Beltrami differential has the correct limits according to figure \ref{fig:Ped}. For $\Lambda\rightarrow0$, it is immediate that we obtain the twistor space of self-dual Taub-NUT.
For $\Lambda\rightarrow\infty$ and $\nu^2\rightarrow0$ with $\Lambda\nu^2$ fixed, we obtain the holomorphic coordinates of Eguchi-Hanson space \cite{Bittleston:2023bzp}.% space with the role of $\lambda_\alpha$ and $\mu^{\da}$ exchanged which was expected from the conformal inversion in equation \eqref{eq:conf}. %This can already be seen on the level of the Beltrami differential in \eqref{eq:NablaQTN} where the first term is the Beltrami differential of self-dual Taub-NUT and the second term is that of Eguchi-Hanson space, up to a conformal inversion.

%The Pedersen twistor space \cite{pedersen1985geometry,Pedersen:1986vup} is given by a line bundle over $Q_{\Lambda/4M}$ with a transition function which is closely related to \eqref{eq:TransitionFunctionPhi}. We conjecture that our deformed twistor space agrees with the Pedersen twistor space and that the missing factor in the transition function comes from the fact that the charts $\lambda_0\neq 0$ and $\lambda_1 \neq 0$ are not the appropriate charts over $Q_{\lambda/4M}$. For details on appropriate charts we refer to \cite{Pedersen:1986vup, pedersen1985geometry} and the companion paper. Note that equation \eqref{eq:TransitionFunctionPhi} reduces to the exponential transition function of equation \eqref{eq:TransitionSDTN} in the limit $\Lambda\rightarrow 0$.  Moreover, in the limit $\Lambda\rightarrow\infty$ and $M\rightarrow\infty$ with $\Lambda/M$ fixed, we obtain the holomorphic coordinates of Eguchi-Hanson space with the role of $\lambda_\alpha$ and $\mu^{\da}$ exchanged which was expected from the conformal inversion in equation \eqref{eq:conf}. This can already be seen on the level of the Beltrami differential in \eqref{eq:NablaQTN} where the first term is the Beltrami differential of self-dual Taub-NUT and the second term is that of Eguchi-Hanson space, up to a conformal inversion.

\subsection{Celestial symmetries from twistor space}
\label{subsec:CelSymmTwistor}
Beyond viewing the twistor space of Eguchi-Hanson space as a backreaction, \cite{Bittleston:2022nfr} obtained the celestial chiral algebra of self-dual gravity from considering the action of $\{\,\,,\,\,\}_0$ on generators built out of the deformed holomorphic coordinates. We will follow the same strategy but since $\Lambda\neq 0$ we will have to use $\{\,\,,\,\,\}_\Lambda$. Acting with $\{\,\,,\,\,\}_\Lambda$ on generators in the undeformed holomorphic coordinates, led to the recent cosmological constant deformation of $L\mathfrak{ham}(\mathbb{C}^2)$ \cite{Bittleston:2024rqe, Taylor:2023ajd, Alday:2023jdk}. In the supplemental material section \ref{sec:Bracket}, we derive the action of $\{\,\,,\,\,\}_\Lambda$ on any pair of holomorphic coordinates  $\phi^\pm,\eta, X,Y,Z$ leading to the main result in equation \eqref{eq:PoissonQTN2}.
This can be used to compute the bracket of two generators $(\phi^+)^a(\phi^-)^b \eta^c X^d Y^e Z^f$, where $a,b,c \in \mathbb{N}$ and $d,e,f \in \mathbb{Z}$, which is the natural generalisation of \cite{Bittleston:2024rqe,Adamo:2021lrv, Bittleston:2023bzp}. We still need to impose the right scaling by demanding that the generators have weight $2$, i.e. $a+b+\tfrac{1}{2}(c+d+e+f)=2$, and impose relations between the different coordinates. For the latter, we note that the subspaces generated by
\bea
\label{eq:Poissonideals}
\langle \{ \phi^+ \phi^- -\eta\}\rangle\,,\qquad
\langle \{XY-Z^2+\tfrac{\Lambda\nu^2}{2}\eta^2\}\rangle\,,
\eea
are Poisson ideals, which means we can use these relations to only consider generators of the form
\bea
\label{eq:reducedgenerators}
&w[p,q,2i,2j]=(\phi^+)^p(\phi^-)^q X^i Y^j\\
&w[p,q,2i+1,2j+1]=(\phi^+)^p(\phi^-)^q X^i Y^jZ\,.
\eea
For such $w[p,q,a,b]$ we have the condition $p+q+a+b=2$. Solving for $b=2-p-q-a$ would lead to the standard presentation of generators with $3$ labels but we will not do so to keep certain symmetries between $\mu^\da$ and $\lambda_\alpha$ manifest. The algebra can then be found by explicitly applying the bracket \eqref{eq:PoissonQTN2} to the generators \eqref{eq:reducedgenerators} and using the constraints \eqref{eq:Poissonideals} which has been spelled out in the supplemental material section \ref{sec:Commutators}.
For instance, considering two generators of the first kind leads to
\begin{widetext}
\bea
\label{eq:QTNAlg1}
\{w[p,q,2i,2j],w[r,s,2k,2l]\}_\Lambda &=(ps-qr)w[p+r-1,q+s-1,2(i+k),2(j+l)]\\
&+4\Lambda(il-jk) w[p+r,q+s,2(i+k-1)+1,2(j+l-1)+1]\\
&-\frac{\Lambda\nu^2}{2}((p-q)(k+l)-(r-s)(i+j))w[p+r+1,q+s+1,2(i+k-1),2(j+l-1)]\,,
\eea
\end{widetext}
Although \eqref{eq:QTNAlg1} and more generally \eqref{eq:QTNAlg} look fairly involved, it is the unique algebra, that respects the expected symmetries, obeys the Jacobi identity and reduces to the known algebras in the limits of section \ref{subsec:CelSymmTwistor}. Indeed, when transformed to the slightly different coordinates of \cite{Bittleston:2024rqe} it can be seen to be a further deformation of the  $\Lambda$-deformed algebra considered in \cite{Bittleston:2024rqe, Taylor:2023ajd}. In fact, we obtain a $\mathbb{Z}_2$-quotient of the AdS$_4$ algebra in the limit $M\rightarrow\infty$, which is an expected feature due to the coordinates \eqref{eq:XYZ} in analogy to \cite{Bittleston:2024rqe}. %The other two types of commutators are displayed in section \ref{sec:Commutators} and it has been explicitly checked that the resulting algebra obeys the Jacobi identity. 
After rescaling the generators by $\Lambda$, related to the prefactor of \eqref{eq:EH}, and sending $\Lambda\rightarrow\infty$ and $M\rightarrow\infty$, keeping $c^2=\tfrac{\Lambda\nu^2}{2}$ constant, the algebra becomes the expected algebra of Eguchi-Hanson space considered in equation (4.8) of \cite{Bittleston:2022nfr}. The role of $\mu^\da$ and $\lambda_\alpha$ is exchanged as expected from the conformal inversion in equation \eqref{eq:conf}.
In the self-dual Taub-NUT limit, $\Lambda\rightarrow 0$, we find an undeformed $Lw_\wedge$.

\acknowledgments
We thank Tim Adamo, William Biggs, Roland Bittleston, Kasia Budzik, Kevin Costello, Adam Kmec, Lionel Mason, Chris Pope, Maria Rodriguez, Sean Seet, Atul Sharma, David Skinner and Andrew Strominger for helpful discussions. Moreover, we thank Atul Sharma for encouragement and comments on the draft. SH also thanks the organisers of the '2024 summer school on celestial holography' and the hospitality of Perimeter Institute and Roland Bittleston. GB is supported by a joint Clarendon Fund and Merton College Mathematics Scholarship. SH is supported by the STFC HEP Theory Consolidated grant ST/T000694/1 and St. John's College, Cambridge.

%%%%%%%%%%%%%%%%%%%%%%%%%%%%%%%%%%%%%%%%%%%

\bibliography{references}
\bibliographystyle{apsrev4-1}

%%%%%%%%%%%%%%%%%%%%%%%%%%%%%%%%%%%%%%%%%%%

\clearpage
\widetext
\begin{center}
    \textbf{Supplemental material}
\end{center}
In the remaining part of this work, we provide some of the technical material relating the various spacetime metrics to each other, deriving holomorphic coordinates and the form of the deformed Jacobi bracket.

%%%%%%%%%%%%%%%%%%%%%%%%%%%%%%%%%%%%%%%%%%%

\section{More on the the Pedersen metric}
\label{sec:Diffeos}
\subsection{From the Pleba\'{n}ski-Demia\'{n}ski to the Pedersen metric}
Let us first derive the triaxial Pedersen metric \eqref{eq:PedMet} from the Pleba\'{n}ski-Demia\'{n}ski metric \eqref{eq:TNAdS4}. At the self-dual point \eqref{eq:self-dualpoint} with $a=0$ and after the Wick rotation $t\mapsto \im t$, the Pleba\'{n}ski-Demia\'{n}ski metric is
\begin{equation}
    \d s^2=\frac{1}{U(r)}(\d t-2M\cos\theta\,\d\phi)^2+U(r)\d r^2+(r^2-M^2)(\d \theta^2+\sin^2\theta\,\d\phi^2)\,,\label{eq:SDTNADS4}
\end{equation}
where $r\geq M$, $(\theta,\phi)$ are coordinates on $S^2$ and $t\sim t+8\pi M$ is a coordinate on $S^1$. We have also introduced
\begin{equation}
    U(r)=\frac{r+M}{r-M}\left(1+\frac{(r-M)(r+3M)}{l^2}\right)^{-1}\,.
\end{equation}
Performing the diffeomorphism
\begin{equation}
    \rho=\sqrt{2Ml^2\frac{r-M}{l^2+2M(r-M)}}\,,\qquad \psi=\frac{t}{2M}\,,
\end{equation}
and introducing 
\begin{equation}
    \nu^2=\frac{1}{4M^2}-\frac{1}{l^2}\,,
\end{equation}
brings the metric into the form
\begin{equation}
    \d s^2=\frac{1}{(1-\rho^2/l^2)^2}\left[4\frac{1+\nu^2\rho^2}{1+\nu^2\rho^4/l^2}\d\rho^2+\rho^2(1+\nu^2\rho^2)(\d \theta^2+\sin^2\theta\,\d\phi^2)+\rho^2\frac{1+\nu^2\rho^4/l^2}{1+\nu^2\rho^2}(\d \psi-\cos\theta\,\d\phi)^2\right]\,,\label{eq:PedMet_with_rho}
\end{equation}
where $0\leq \rho< l$ and $0\leq \psi<4\pi$. This is precisely the metric \eqref{eq:PedMet}, after relabelling $\rho\mapsto r$ and introducing the 1-forms
\begin{equation}
    \sigma_1=\frac{1}{2}(\cos\psi\,\d\theta+\sin\psi\,\sin\theta\,\d\phi)\,,\qquad\sigma_2=\frac{1}{2}(-\sin\psi\,\d\theta+\cos\psi\,\sin\theta\,\d\phi)\,,\qquad \sigma_3=\frac{1}{2}(\d\psi-\cos\theta\,\d\phi)\,.
\end{equation}

In order to obtain the generalized Gibbons-Hawking form of the metric, consider the further diffeomorphism
\begin{equation}
    \rho=\sqrt{\frac{l}{\nu}\tan\chi}\,.
\end{equation}
This brings the metric into the metric \eqref{eq:s3_gibbons-hawking}. Similarly, if we start from the oblate metric \eqref{eq:PedMet_with_rho} and consider the analytic continuation $\nu\mapsto\im\nu$, we find the prolate metric \eqref{eq:eq:h3_gibbons-hawking} using the diffeomorphism
\begin{equation}
    \rho=\sqrt{\frac{l}{\nu}\tanh\chi}\,.
\end{equation}
\subsection{Various limits of the Pedersen metric}
Let us now discuss the various limits of the Pedersen metric that were previously studied in the context of celestial holography \cite{Crawley:2021auj, Crawley:2023brz, Bittleston:2023bzp, Taylor:2023ajd, Costello:2023hmi, Bittleston:2024efo} and explain Figure \ref{fig:Ped}.
\paragraph{Self-dual Taub-NUT}
\label{para:SDTNmetric}
We first consider the limit in which the cosmological constant is vanishing, $l\to\infty$. It is easier to understand this limit from the form \eqref{eq:SDTNADS4} of the Pedersen metric, which reduces to
\bea
\d s^2&=\left(1+\frac{2M}{r}\right)^{-1}(\d t+2M\,\cos\theta\,\d\phi)^2+\left(1+\frac{2M}{r}\right)(\d r^2+r^2\d\Omega_2^2)\,,
\eea
after a further shifting $r\mapsto r+M$. This is precisely self-dual Taub-NUT.

\paragraph{AdS$_4$}
\label{para:AdS4met}
The second natural limit is the limit $\nu\to 0$. In this case, the Pedersen metric in the form \eqref{eq:PedMet} is
\begin{equation}
    \d s^2=\frac{4}{(1-r^2/l^2)^2}(\d r^2+r^2(\sigma_1^2+\sigma_2^2+\sigma_3^2))\,.
\end{equation}
This is Euclidean AdS$_4$. As expected, the boundary metric reduces to the canonical metric on the 3-sphere parametrized by $(\psi,\theta,\phi)$ \footnote{Be careful not to confuse this 3-sphere with the 3-sphere parametrized by $(\chi,\theta,\phi)$.}.

\paragraph{A singular double cover of Eguchi-Hanson space}
\label{para:EH2met}
In this and the following limit, we focus on the prolate Pedersen metric \eqref{eq:eq:h3_gibbons-hawking}. Consider the limit $\nu,l\to 0$ with $c^2\equiv l/\nu$ held constant and set 
\begin{equation}
\label{eq:conf}
    r=\frac{c^2}{\varrho}\,.
\end{equation}
The metric in triaxial form then becomes
\begin{equation}
\label{eq:EH}
    \d s^2=\frac{\d \varrho^2}{1-c^4/\varrho^4}+\varrho^2(\sigma_1^2+\sigma_2^2)+\varrho^2\left(1-\frac{c^4}{\varrho^4}\right)\sigma_3^2\,,
\end{equation}
up to an overall factor of $l^4/4c^2$. The space-time metric can be extended to the region $\varrho>c$ and is seen to be the Eguchi-Hanson metric. More precisely, the singularity at $\varrho=c$ is not removable without taking a $\Z_2$ quotient, so the spacetime is actually a singular double cover of Eguchi-Hanson space \cite{Eguchi:1978gw}.

\paragraph{Burns space and $\widetilde{\mathbb{CP}}^2$}
\label{para:Burns}
Setting $\nu l=1$ in the prolate Pedersen metric \eqref{eq:eq:h3_gibbons-hawking}, we recover the Burns space metric \cite{Costello:2023hmi} up to a conformal prefactor. In fact, performing the change of coordinates
\begin{equation}
    r^2=\frac{R^2}{2-R^2/l^2}\,,
\end{equation}
in the triaxial form of the metric, we find
\begin{equation}
    \d s^2=\frac{2}{(1-R^2/l^2)^2}(\d R^2+\rho^2\sigma_3^2)+\frac{2}{1-R^2/l^2}(\sigma_1^2+\sigma_2^2)\,,
\end{equation}
which is the Fubini-Study metric on a non-compact version of $\CP^2$ commonly referrend to as Bergmann space and denoted by $\widetilde{\mathbb{CP}}^2$. As a homogeneous space it is described by $\widetilde{\mathbb{CP}}^2=SU(2,1)/U(2)$ and it is the non-compact dual of $\mathbb{CP}^2=SU(3)/U(2)$ in the sense of \cite{Helgason:1968bvw}. Both $\widetilde{\mathbb{CP}}^2$ and $\mathbb{CP}^2$ with a point removed are conformally equivalent to Burns space \cite{Costello:2023hmi} and \eqref{eq:ProlPedMetconf} gives the corresponding conformal factor. %Note that the oblate Pedersen metric is generically a complete metric on a $4$-dimensional ball. This changes as we approach $\nu l=1$. $\widetilde{\mathbb{CP}}^2$ does not have the same topology as the $4$-dimensional ball, by having a non-trivial $2$-cycle. In fact, we only find the Fubini-Study metric an open subset of $\widetilde{\mathbb{CP}}^2$ which then can then be extended to all of $\widetilde{\mathbb{CP}}^2$ including this $2$-cycle. Also 
Note that $\widetilde{\mathbb{CP}}^2$ has a degenerate boundary conformal structure since the prolate Pedersen metric generically has the boundary conformal structure
\begin{equation}
\label{eq:prolatesquashedS3}
    \d s^2_{3}=\sigma_1^2+\sigma_2^2+\frac{1}{1-\nu^2l^2}\sigma_3^2\,,
\end{equation}
and we set $\nu l=1$. The similarity to the boundary of Minkowski space has been previously commented on in \cite{Britto-Pacumio:1999dwj}.

%%%%%%%%%%%%%%%%%%%%%%%%%%%%%%%%%%%%%%%%%%%

\section{The curved twistor space and celestial symmetries}
\subsection{Holomorphicity of the deformed coordinates}
\label{sec:Holom}
Let us check explicitly that the coordinates discussed in section \ref{sec:sec3} are holomorphic for $\bar{\nabla}_\Lambda$ of equation \eqref{eq:NablaQTN}. First, we have 
\begin{equation}
Y^{\alpha\beta}=\lambda^\alpha\lambda^\beta-\frac{\Lambda\nu^2 \eta^2}{2}\frac{\hat\lambda^\alpha\hat\lambda^\beta}{\la\lambda\,\hat\lambda\ra^2}\,.
\end{equation} 
which obeys
\begin{equation}
    \begin{aligned}
        \bar\nabla_\Lambda Y^{\alpha\beta}&=2\lambda^{(\alpha}\bar\nabla_\Lambda\lambda^{\beta)}-2\frac{\Lambda\nu^2 \eta^2}{2}\frac{\hat\lambda^{(\alpha}}{\la\lambda\,\hat\lambda\ra}\bar\nabla_\Lambda\frac{\hat\lambda^{\beta)}}{\la\lambda\,\hat\lambda\ra}\\&=2\frac{\Lambda\nu^2 \eta^2}{2}\bar e^0\frac{\lambda^{(\alpha}\hat\lambda^{\beta)}}{\la\lambda\,\hat\lambda\ra}-2\frac{\Lambda\nu^2 \eta^2}{2}\bar e^0 \frac{\hat\lambda^{(\alpha}\lambda^{\beta)}}{\la\lambda\,\hat\lambda\ra}\\&=0\,.
    \end{aligned}
\end{equation}
$\eta=\mu^+\mu^-$ is manifestly holomorphic since $h$ only depends on $\mu^{\dot{\alpha}}$ through $\eta$ and $\{\eta, \eta\}_\Lambda=0$. For $\phi^+$, we can easily see that 
\begin{equation}
    \begin{aligned}
        \bar\nabla_\Lambda\phi^+%&=%(\bar\nabla_\Lambda\mu^{\dot 0})\exp\left(\frac{1}{2\sqrt{4\Lambda M}}\log\frac{1+\sqrt{4\Lambda M}\dfrac{\eta f(\lambda)}{4M}}{1-\sqrt{4\Lambda M}\dfrac{\eta f(\lambda)}{4M}}\right)+\mu^{\dot 0}\bar\nabla_\Lambda\exp\left(\frac{1}{2\sqrt{4\Lambda M}}\log\frac{1+\sqrt{4\Lambda M}\dfrac{\eta f(\lambda)}{4M}}{1-\sqrt{4\Lambda M}\dfrac{\eta f(\lambda)}{4M}}\right)\\&=\phi^+\left(-\frac{\eta\bar e^0}{4M}+\frac{1}{2\sqrt{4\Lambda M}}\bar\nabla_\Lambda\log\frac{1+\sqrt{4\Lambda M}\dfrac{\eta f(\lambda)}{4M}}{1-\sqrt{4\Lambda M}\dfrac{\eta f(\lambda)}{4M}}\right)\\
        &=\frac{\nu^2\eta\phi^+}{2}\left(-\bar e^0+\frac{1}{1-\dfrac{\Lambda\nu^2\eta^2 f(\lambda)^2}{2}}\bar\nabla_\Lambda f(\lambda)\right)
        =0\,,
    \end{aligned}
    \end{equation}
where we used that $\frac{\hat\lambda_\alpha}{\la\lambda\,\hat\lambda\ra}\frac{\p f}{\p\lambda_\alpha}=-f^2$ holds in both patches. Similarly $\bar{\nabla}_\Lambda\phi^-=0$.

\subsection{Jacobi bracket in deformed coordinates}
\label{sec:Bracket}

Let us now derive the action of $\{\,\,,\,\,\}_\Lambda$ on any pair of holomorphic coordinates  $\phi^\pm,\eta, X,Y,Z$. We derive it in part by acting on holomorphic coordinates with
\bea
\label{eq:LambdaBrack2}
&\{f,g\}_\Lambda=\varepsilon^{\dot\alpha\dot\beta}\frac{\p f}{\p\mu^{\dot\alpha}}\frac{\p g}{\p\mu^{\dot\beta}}+\Lambda\,\varepsilon_{\alpha\beta}\frac{\p f}{\p\lambda_\alpha}\frac{\p g}{\p\lambda_\beta}\,,
\eea
and in part from the two consistency conditions that the Jacobi identity needs to hold and that the subspaces generated by \eqref{eq:Poissonideals}
need to be Poisson-ideals. 

%In this section we will discuss our conjectured Poisson bracket on the curved twistor space (more precisely, the Jacobi bracket when working on projective twistor space). We derive it in part by acting on holomorphic coordinates with
%\bea
%\label{eq:LambdaBrack2}
%&\{f,g\}_\Lambda=\varepsilon^{\dot\alpha\dot\beta}\frac{\p f}{\p\mu^{\dot\alpha}}\frac{\p g}{\p\mu^{\dot\beta}}+\Lambda\,\varepsilon_{\alpha\beta}\frac{\p f}{\p\lambda_\alpha}\frac{\p g}{\p\lambda_\beta}\,,
%\eea
%and in part from the two consistency conditions that the Jacobi identity needs to hold and the subspaces \eqref{eq:Poissonideals} need to be Poisson ideals. 
First, we act with \eqref{eq:LambdaBrack} on a pair of $Y_{\alpha \beta}$ giving
\begin{equation}
    \{Y_{\alpha\beta}, Y_{\gamma\delta}\}_\Lambda=2\Lambda(\varepsilon_{\beta\gamma}Y_{\alpha\delta}+\varepsilon_{\alpha\delta}Y_{\beta\gamma})\,.
\end{equation}
This is equivalent to $X,Y,Z$ obeying the defining relations of $\mathfrak{sl}_2$ similar to \cite{Bittleston:2023bzp} \be \label{eq:Poisson-bracketXYZ}  
 \{X,Y\}_\Lambda = 4\Lambda Z\,,\qquad \{X,Z\}_\Lambda = 2\Lambda X\,,\qquad \{Y,Z\}_\Lambda = -2 \Lambda Y\,.
\ee
Moreover, using \eqref{eq:LambdaBrack2} we can immediately see 
\bea
\{\phi^{\pm}, \eta\}_\Lambda=\pm \phi^{\pm} \,,\qquad 
\{Y_{\alpha \beta},\eta \}_\Lambda=0\,,\qquad 
\{\phi^+,\phi^-\}_\Lambda=\{\mu^+,\mu^-\}_\Lambda=1\,.
\eea
Demanding now that the subspace generated by $\langle \{XY-Z^2+\frac{\Lambda\nu^2}{2}\eta^2\}\rangle$ is a Poisson ideal and the resulting bracket has to obey the Jacobi identity uniquely fixes the remaining brackets (under the additional assumption that $X$ and $Y$ should be treated symmetrically) to be
\be
\label{eq:magicbracket}
\{\phi^\pm,Y\}_\Lambda=\mp \frac{\Lambda\nu^2}{2}  \phi^\pm \frac{\eta}{X}\,, \qquad \{\phi^\pm,X\}_\Lambda=\mp \frac{\Lambda\nu^2}{2}  \phi^\pm \frac{\eta}{Y}\,,\qquad\{\phi^\pm, Z\}_\Lambda=0\,.
s\ee
We hope to give a more direct derivation of equation \eqref{eq:magicbracket} in the future.

Although a direct derivation is absent, the above discussion leads to strong evidence for the conjectured formula
\bea
\label{eq:PoissonQTN2}
&\{\,\,,\,\,\}_\Lambda= 2\Lambda X \frac{\partial}{\partial X}\wedge\frac{\partial}{\partial Z} -2\Lambda Y \frac{\partial}{\partial Y}\wedge\frac{\partial}{\partial Z} +4\Lambda Z\frac{\partial}{\partial X}\wedge\frac{\partial}{\partial Y}+\frac{\partial}{\partial \phi^+}\wedge\frac{\partial}{\partial \phi^-}
+\phi^-\frac{\partial}{\partial  \phi^+}\wedge\frac{\partial}{\partial \eta}-\phi^+ \frac{\partial}{\partial \phi^-}\wedge\frac{\partial}{\partial \eta}\\
&+\frac{\Lambda\nu^2}{2}\,\eta \Bigg(\frac{\phi^-}{X}\frac{\partial}{\partial \phi^-}\wedge\frac{\partial}{\partial Y} +\frac{\phi^-}{Y}\frac{\partial}{\partial \phi^-}\wedge\frac{\partial}{\partial X}-\frac{\phi^+}{X}\frac{\partial}{\partial \phi^+}\wedge\frac{\partial}{\partial Y} -\frac{\phi^+}{Y}\frac{\partial}{\partial \phi^+}\wedge\frac{\partial}{\partial X}\Bigg)\,.
\eea
In section \ref{subsec:CelSymmTwistor}, we discussed that the correct limits are obtained from equation \eqref{eq:PoissonQTN2} according to figure \ref{fig:Ped}.

\subsection{Deriving the $2$-parameter deformation of $Lw_\wedge$}
\label{sec:Commutators}

Equation \eqref{eq:PoissonQTN2} can be used to compute the bracket of two generators 
\begin{equation}
(\phi^+)^a(\phi^-)^b \eta^c X^d Y^e Z^f\,,
\end{equation}
where $a,b,c \in \mathbb{N}$ and $d,e,f \in \mathbb{Z}$. Since $\phi^\pm, \eta$ are deformations of the $\mu^\da$ coordinates and $X,Y,Z$ are deformations of the $\lambda_\alpha$ coordinates, these generators and their index ranges are the natural generalisation of the treatment in the literature \cite{Bittleston:2024rqe,Adamo:2021lrv, Bittleston:2023bzp}. We still need to impose the right scaling by demanding that the generators have weight $2$, i.e. $a+b+\tfrac{1}{2}(c+d+e+f)=2$, and impose the relations \eqref{eq:Poissonideals} between the different coordinates. Let us initially only work on the support of $\phi^+\phi^-=\eta$ with no relation between $X,Y,Z,\eta$ imposed. A generator of weight $2$ is then given by
\begin{equation}
w[a,b,c,d,e]=(\phi^+)^a (\phi^-)^bX^cY^dZ^e\,,
\end{equation}
where $a,b\in\mathbb{N}_0$,  $c,d,e \in \mathbb{Z}$ and $a+b+\tfrac{1}{2}(c+d+e)=2$. Using \eqref{eq:PoissonQTN2}, leads to the Lie-algebra
\begin{equation}
\begin{aligned}
\label{eq:QTNbeforeQuotient}
&\{w[p,q,i,j,k],w[r,s,l,m,n]\}_\Lambda=(ps-qr)\, w[p+r-1,q+s-1,i+l,j+m,k+n]\\
&+2\Lambda\bigg(((i-j)n-(l-m)k)\,w[p+r,q+s,i+l,j+m,k+n-1]
\\
&+2(im-jl)\,w[p+r,q+s,i+l-1,j+m-1,k+n+1]\bigg)\\
&- \frac{\Lambda\nu^2}{2}((p-q)(l+m)-(r-s)(i+j))\,w[p+r+1,q+s+1,i+l-1,j+m-1,k+n]\,,
\end{aligned}
\end{equation}
which has been explicitly checked to obey the Jacobi identity. Imposing the further constraint $Z^2=XY+\tfrac{\Lambda\nu^2}{2}\eta^2$ means that we can always solve for $X,Y,\eta$ whenever we have $2$ or more powers of $Z$. This leaves us with the generators \eqref{eq:reducedgenerators}.  

Using the relation  $XY-Z^2=-\tfrac{\Lambda\nu^2}{2}\eta^2$ in  \eqref{eq:QTNbeforeQuotient} then leads us to the final algebra. %To not print this final algebra twice, we immediately include the central extension discussed in section \ref{subsec:CelSymmTwistor}.
\begin{equation}
\label{eq:QTNAlg}
\begin{aligned}
&\{w[p,q,2i,2j],w[r,s,2k,2l]\}_\Lambda =(ps-qr)w[p+r-1,q+s-1,2(i+k),2(j+l)]\\
&+4\Lambda(il-jk) w[p+r,q+s,2(i+k-1)+1,2(j+l-1)+1]\\
&-\frac{\Lambda\nu^2}{2}((p-q)(k+l)-(r-s)(i+j))w[p+r+1,q+s+1,2(i+k-1),2(j+l-1)]
\\
&\{w[p,q,2i,2j],w[r,s,2k+1,2l+1]\}_\Lambda=(ps-qr)w[p+r-1,q+s-1,2(i+k)+1,2(j+l)+1]\\
&+2\Lambda(i(2l+1)-j(2k+1)) w[p+r,q+s,2(i+k),2(j+l)]\\
&+2\Lambda^2\nu^2(il-jk) w[p+r+2,q+s+2,2(i+k-1),2(j+l-1)]\\
&-\frac{\Lambda\nu^2}{2}((p-q)(k+l)-(r-s)(i+j))w[p+r+1,q+s+1,2(i+k-1)+1,2(j+l-1)+1]\\
&\{w[p,q,2i+1,2j+1],w[r,s,2k+1,2l+1]\}_\Lambda\\
&=(ps-qr)w[p+r-1,q+s-1,2(i+k+1),2(j+l+1)]\\
&+\frac{\Lambda\nu^2}{2}(ps-qr)w[p+r+2,q+s+2,2(i+k),2(j+l)]\\
&+\Lambda((2i+1)(2l+1)-(2j+1)(2k+1)) w[p+r,q+s,2(i+k)+1,2(j+l)+1]\\
&+ 2\Lambda^2\nu^2(il-jk) w[p+r+2,q+s+2,2(i+k-1)+1,2(j+l-1)+1]\\
&-\frac{\Lambda\nu^2}{2}((p-q)(k+l)-(r-s)(i+j))w[p+r+1,q+s+1,2(i+k),2(j+l)]\\
&-\Big(\frac{\Lambda\nu^2}{2}\Big)^2((p-q)(k+l)-(r-s)(i+j))w[p+r+3,q+s+3,2(i+k-1),2(j+l-1)]\,.
\end{aligned}
\end{equation}
 This is the unique Lie-algebra, that respects the expected symmetries, obeys the Jacobi identity and reduces to the known algebras in certain limits as discussed in section \ref{subsec:CelSymmTwistor}.   

\end{document}